\newcommand{\BCSGO} {BaCu$_2$(Si$_{1-x}$Ge$_{x}$)$_2$O$_7$\xspace}
\newcommand{\musr}{$\mu$SR\xspace}
\newcommand{\TN}{$\mathrm{T_N}$\xspace}
\newcommand{\mo}{m$_0$\xspace}

\documentclass[prb,twocolumn,floatfix,preprintnumbers,amsmath,amssymb,superscriptaddress]{revtex4}
\usepackage{graphicx}
\usepackage{dcolumn}
\usepackage{bm}
\usepackage{color}
\usepackage{amsmath}
\usepackage{xspace}

\begin{document}

\title{Inhomogeneous ordering in weakly coupled Heisenberg $S=1/2$ chains with random bonds}
\author{M. Thede}
\email{mthede@phys.ethz.ch}
\affiliation{Neutron Scattering and Magnetism, Laboratory for Solid State
Physics, ETH Z\"urich, Z\"urich, Switzerland}
\affiliation{Laboratory for Muon Spin Spectroscopy, Paul Scherrer Institut, CH-5232 Villigen PSI, Switzerland}
\author{T. Haku}
\affiliation{Neutron Scattering Laboratory, Institute for Solid State Physics, University of Tokyo, Tokai, Ibaraki, 319-1106, Japan}
\author{T. Masuda}
\affiliation{Neutron Scattering Laboratory, Institute for Solid State Physics, University of Tokyo, Tokai, Ibaraki, 319-1106, Japan}
\author{C. Baines}
\affiliation{Laboratory for Muon Spin Spectroscopy, Paul Scherrer Institut, CH-5232 Villigen PSI, Switzerland}
\author{E. Pomjakushina}
\affiliation{Laboratory for Developments and Methods, Paul Scherrer Institute, CH-5232 Villigen PSI, Switzerland}
\author{G. Dhalenne}
\affiliation{Synth\`{e}se, Propri\'{e}t\'{e}s et Mod\'{e}lisation des Mat\'{e}riaux, Universite Paris-Sud, 91405 Orsay cedex, France.}
\author{A. Revcolevschi}
\affiliation{Synth\`{e}se, Propri\'{e}t\'{e}s et Mod\'{e}lisation des Mat\'{e}riaux, Universite Paris-Sud, 91405 Orsay cedex, France.}
\author{E. Morenzoni}
\affiliation{Laboratory for Muon Spin Spectroscopy, Paul Scherrer Institut, CH-5232 Villigen PSI, Switzerland}
\author{A. Zheludev}
 \email{zhelud@ethz.ch}
 \homepage{http://www.neutron.ethz.ch/}
\affiliation{Neutron Scattering and Magnetism, Laboratory for Solid State
Physics, ETH Z\"urich, Z\"urich, Switzerland}

\begin{abstract}
Long range magnetic ordering in the quasi-one-dimensional random-bond antiferromagnet \BCSGO is studied in \musr experiments as a function of disorder strength. Compared to the disorder-free parent materials, the saturation ordered moment is found to be considerably reduced. Moreover, even in weakly disordered species, the magnetically ordered state is shown to be highly inhomogeneous. The results are interpreted in terms of weakly coupled random spin chains, governed by the ``infinite randomness`` fixed point.
\end{abstract}

\maketitle

\section{Introduction}
Arbitrary weak disorder can qualitatively alter one-dimensional spin systems. A case in point is the Heisenberg $S=1/2$ antiferromagnetic chain with random bond strengths. Even though all exchange constants remain antiferromagnetic, the ground state and low energy excitations are quite distinct from those in Bethe's uniform spin chain.\cite{Bethe1931} Regardless of the details of disorder, the system  flows to the ``infinite randomness'' fixed point, represented by the so-called Random Singlet (RS) phase.\cite{Ma1979,Dasgupta1980,Fisher1994,Motrunich2000} In the latter, all strong bonds are eliminated via a decimation procedure. What remains is a collection of non-interacting {\it effective} $S=1/2$ antiferromagnetic dimers. The distribution of dimer strengths diverges at $\omega\rightarrow 0$ and is universal in this limit. As a result, and all low-temperature thermodynamic properties and correlation functions acquire {\it universal} scaling forms.

\begin{figure}[h]
 \includegraphics[width=\columnwidth]{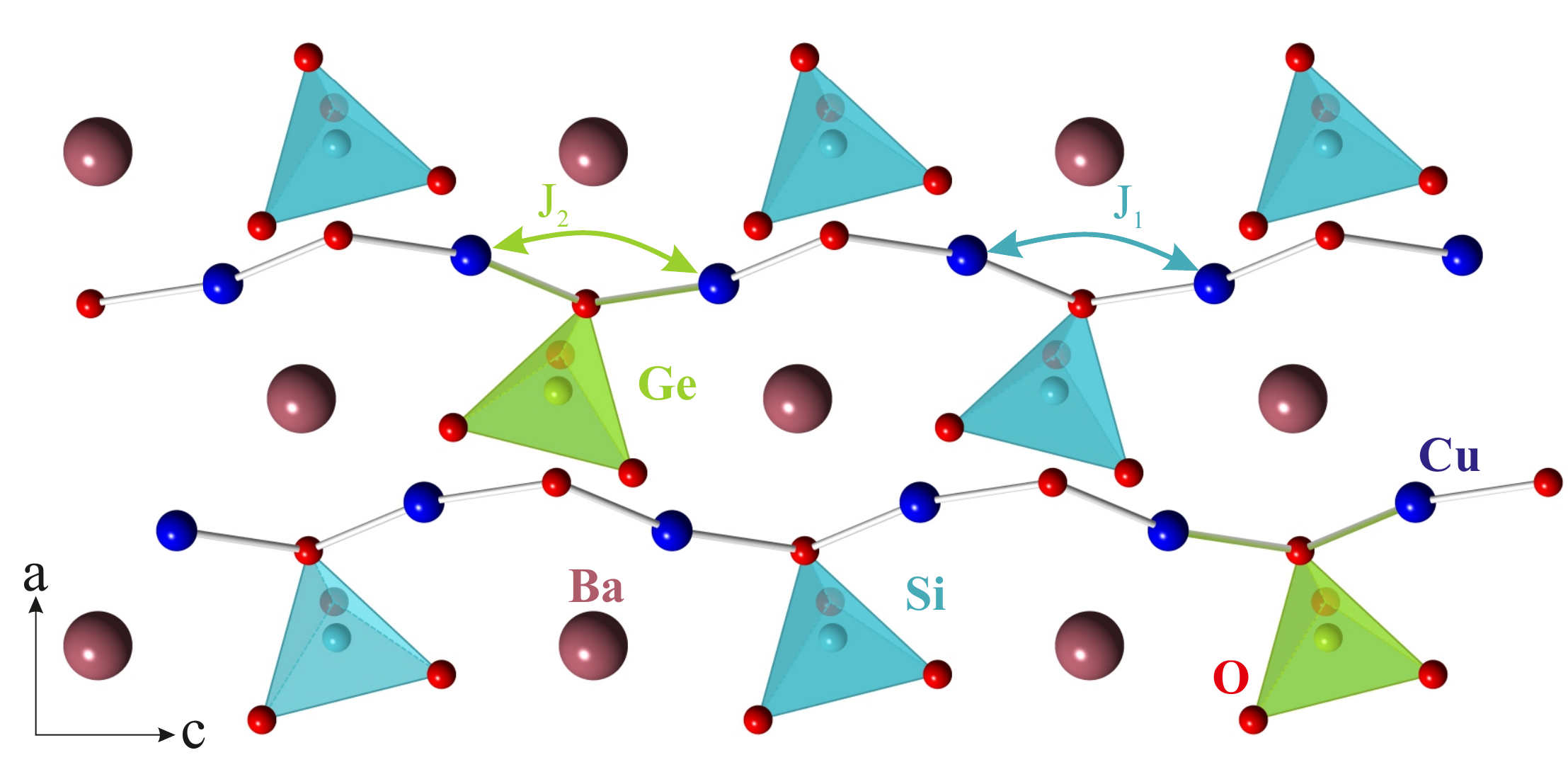}
 \caption{\label{struc} (Color online) Schematic view of the crystal structure of BaCu$_2$(Si$_{1-x}$Ge$_x$)$_2$O$_7$, emphasizing the Cu-O spin chains.}
 \end{figure}

While the RS state is by now very well understood theoretically, experimental realizations are few. The first was quinolinium-(TCNQ)$_2$, where the pioneering work of Tippie and Clark reported signs of RS physics in bulk properties.\cite{Tippie1981} The only other potential candidate is \BCSGO.\cite{Masuda2004,Masuda2006erratum,Shiroka2011,Casola2012} The disorder-free compounds with $x=0$ and $x=1$ are well-characterized $S=1/2$ chain systems.\cite{Tsukada1999,Zheludev2000-2,ZheludevKenzelmann2001,Kenzelmann2001,Zheludev2002-2} The crystal structure is visualized in Fig.~\ref{struc}.\cite{Oliveira1993} The $S=1/2$-carrying Cu$^{2+}$ are bridged by superexchange interactions via O$^{2-}$ ions, forming chains along the $c$ axis of the orthorhombic structure. The corresponding antiferromagnetic exchange constants are $J=280$~K and $J=540$~K, for $x=0$ and $x=1$, respectively.\cite{Tsukada1999} For the disordered derivatives $0<x<1$, the {\it average} values of $J$ changes almost linearly with $x$.\cite{Yamada2001} However, due to the difference in the ionic radii of Si and Ge, the exchange constants are randomly distributed in the chains. A recent comparison of bulk measurements with numerical simulations has convincingly demonstrated that the distribution of $J$s is, to a  good approximation, a simple bimodal one. In fact, it is just a spatially random distribution of the values in the two parent compounds.\cite{Shiroka2011,Casola2012} The randomness of the distribution reveals itself in an unusual behavior of the low temperature magnetic susceptibility.\cite{Yamada2001,Masuda2004,Shiroka2011} Unfortunately, an early inelastic neutron scattering confirmation of RS physics in \BCSGO \cite{Masuda2004} was later shown to be erroneous.\cite{Masuda2006erratum,Zheludev2007-2} However, recent detailed macroscopic and NMR studies,\cite{Shiroka2011,Casola2012} as well as theoretical work,\cite{Herbrych2013} have firmly confirmed that for $x=0.5$, \BCSGO indeed demonstrated all the key features of the RS state.

Due to weak inter-chain interactions, all \BCSGO materials order magnetically, with N\'{e}el temperatures ranging from 1~K to 10~K.\cite{Tsukada1999,Yamada2001,Kenzelmann2001} Therefore, all previous work aiming to probe RS physics was focused on the paramagnetic phase, to avoid the effects of 3-dimensional order. In contrast, the present paper focuses on the low-temperature ordered phase. We use a range of \musr experiments to demonstrate that even in weakly disordered samples, the 3-dimensionally ordered state is highly unusual. Not only is the ordered moment considerably reduced compared to the parent compounds, it is also highly inhomogeneous. We argue that this inhomogeneity is not a simple reflection of the weak structural disorder, but represents the  intrinsic ``infinitely random'' nature of the RS state in individual chains.

\section{Experimental}
Single crystal samples of \BCSGO with $x=0$, $0.05$, $0.15$, $0.3$, $0.5$, $0.95$ and $1$ were grown using the floating zone method. Some of the crystals were from the batch already used in previous studies \cite{Tsukada1999,Zheludev2000-2,ZheludevKenzelmann2001,Kenzelmann2001,Zheludev2002-2,Masuda2004,Masuda2006erratum,Zheludev2007-2}. Crystallographic characterization structure was performed on a  Bruker AXS single crystal diffractometer with an APEX II CCD detector, with over 3000 Bragg intensities collected for each sample.
Calorimetric measurements were performed with a Quantum Design PPMS in the temperature range 50~mK--250~K. These data were collected on small fragments of masses ranging between 7.5 mg and 22.4 mg. The main focus of this paper is on the \musr experiments. These were performed on the GPS, Dolly and LTF spectrometers at the Swiss Muon Source at the Paul Scherrer Institut. The samples were powders, obtained from parts of the as-grown single crystals. The samples were glued with GE-varnish on a silver plate for LTF, or wrapped in silver polyester foil for the Dolly and GPS experiments. The \musr data were collected in zero field, in weak transversal fields (wTF) of 3~mT or 5~mT, and various longitudinal fields. The spectra were analyzed using \textsc{musrfit},\cite{Suter2012} a tool to analyse time-differential \musr data.

\section{Results and data analysis}

\subsection{Crystallographic characterization}
\begin{figure}[h]
 \includegraphics[width=\columnwidth]{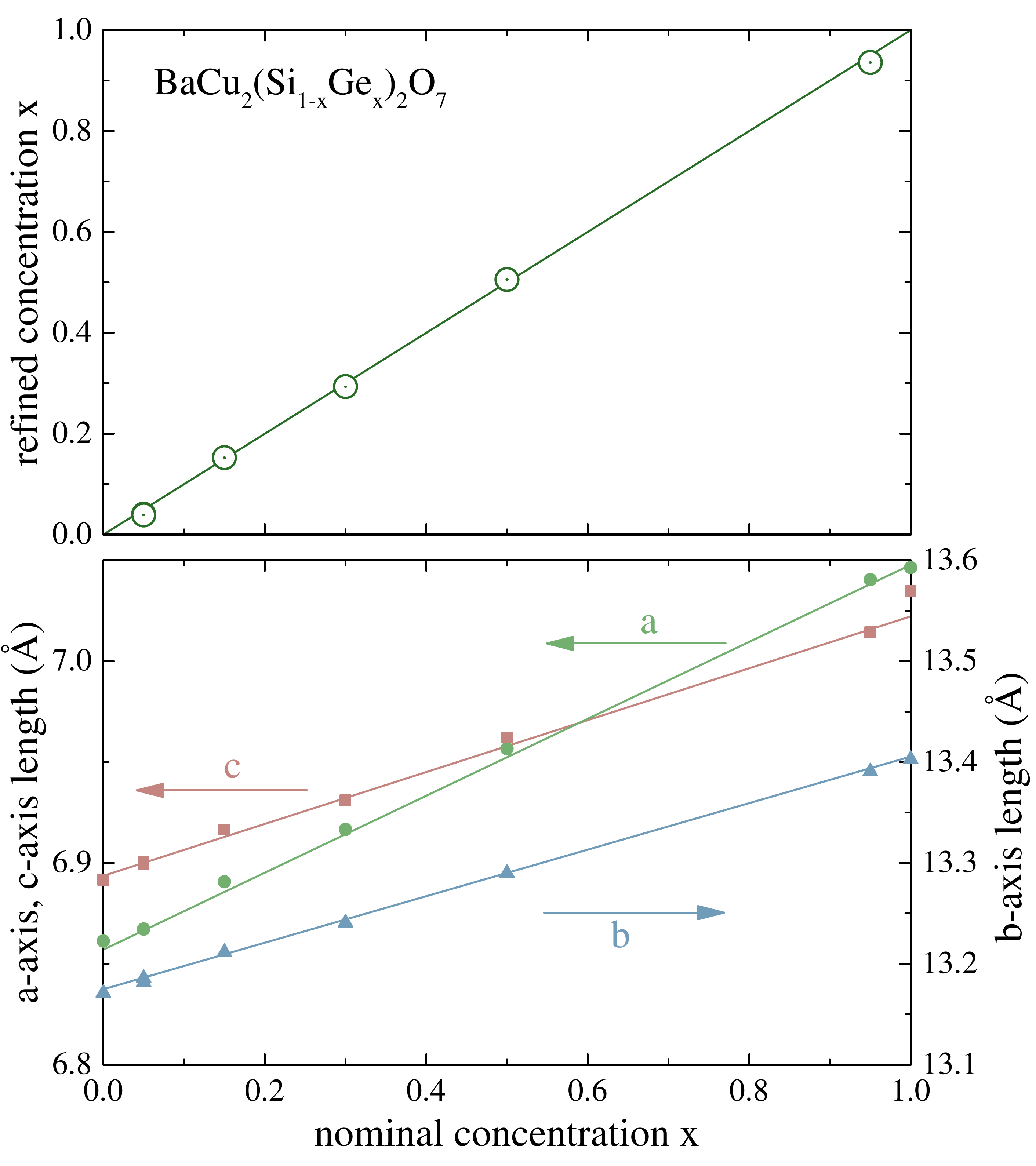}
 \caption{\label{lattice} (Color online)   Top: Ge-concentration determined from single crystal X-ray diffraction for all samples as a function of nominal concentration. Bottom: Lattice parameters as a function of nominal concentration. }
 \end{figure}

In all samples, single crystal X-ray diffraction at room temperature is consistent with the orthorhombic space group \textit{Pnma}. The measured $x$-dependence of the lattice parameters in \BCSGO is monotonic and linear (Fig. \ref{lattice}, bottom), in agreement with previous studies.\cite{Oliveira1993,Yamada2001} A structural refinement confirmed that the actual Ge content in our samples is in excellent agreement with the nominal one (Fig. \ref{lattice}, top). Moreover, small samples taken from different parts of the as-grown crystalline rods showed no detectable composition variation.

\subsection{Specific heat}
 \begin{figure}[h]
 \includegraphics[width=\columnwidth]{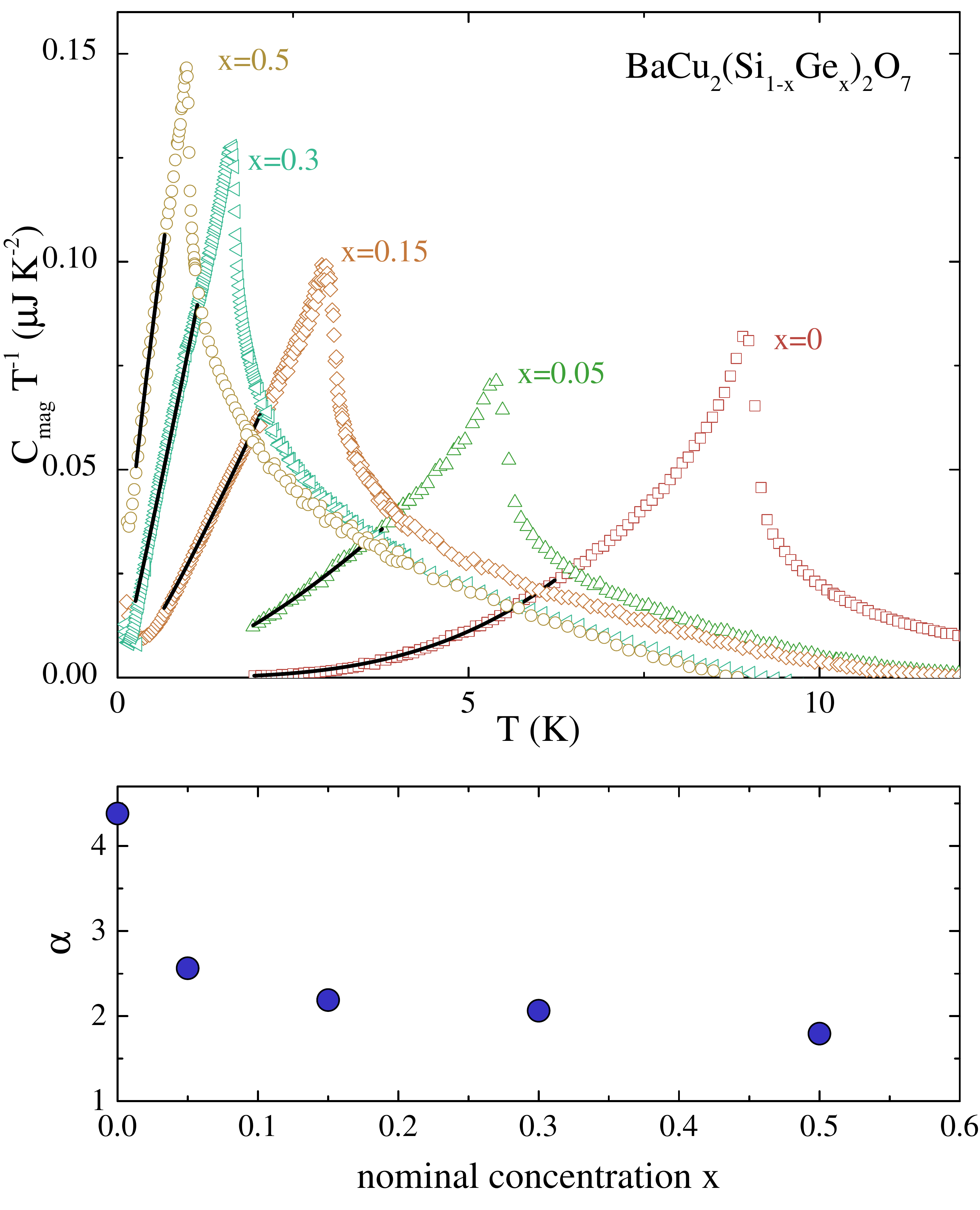}
 \caption{\label{HC} (Color online)  Top: Measured temperature dependence of the specific heat $C/T$ of \BCSGO \space for different concentrations (symbols). Solid lines are power law fits to the data in the ordered state, as described in the text. Bottom: composition dependence of the specific heat power law $C(T)\propto T^\alpha$. }
 \end{figure}
\begin{figure}
 \includegraphics[width=\columnwidth]{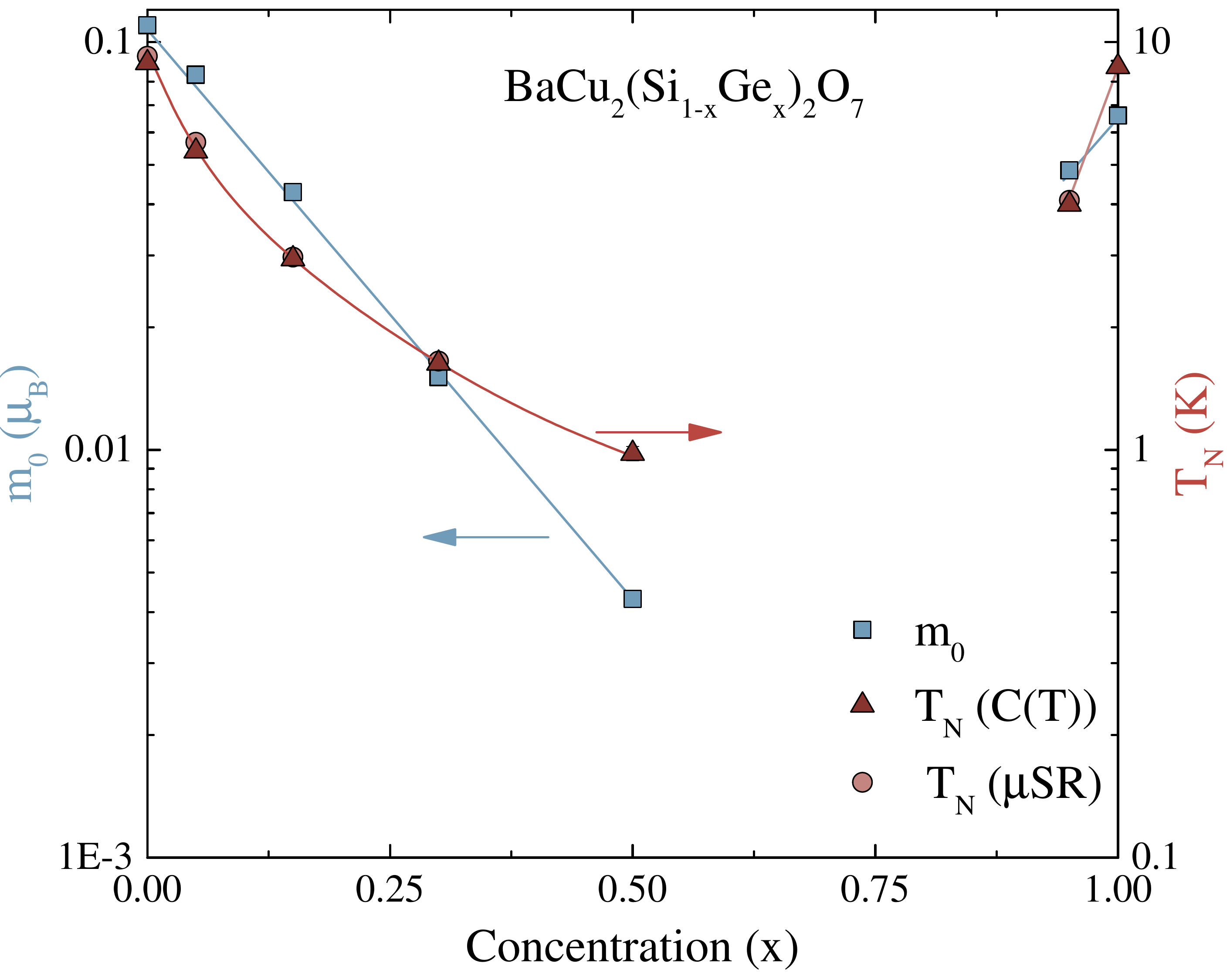}
 \caption{\label{m0} (Color online) Concentration dependence of the ordering temperature \TN \space determined by specific heat and \musr measurements and saturation moment \mo \space.}
 \end{figure}

The main subject of this study being magnetic ordering in \BCSGO, specific heat measurements were used as the most straightforward indicator of the corresponding phase transitions. The focus of these measurements was on Si-rich samples with $x\leq 0.5$. The magnetic contribution C$_{mag}$ was estimated by subtracting a $T^3$ Debye lattice term, extrapolated down to low temperature from the range between 14~K and 25~K. As shown in  Fig.~\ref{HC}, top panel, a pronounced lambda-anomaly is observed in all samples studied. The feature remains sharp and well defined across the entire concentration range. Its position was associated with the N\'{e}el temperature $T_\mathrm{N}$, and is tabulated in Tab. \ref{tab:ordering} and plotted against concentration in Fig.~\ref{m0} (right axis). The two end materials show the maximal $T_\mathrm{N}$, while the lowest value is observed for $x=0.5$. Note that our data on well-characterized single crystals are in contradiction with an earlier work  on polycrystalline samples.\cite{Yamada2001} In the latter, two transitions were claimed for the low concentration regime, and the magnitude of the lambda-anomaly was found to substantially decrease with increasing disorder. We suspect that the discrepancy is due to sample inhomogeneities in the polycrystal experiments.

A striking feature of the measured temperature dependencies of the specific heat is the change in the low-temperature asymptotic in disordered samples, as compared to that in the parent material.
To quantify this behavior, the data in the ranges $0.25\text{~K}<T<0.7~T_\mathrm{N}$ ($x=0.5$, $x=0.3$), $0.65\text{~K}<T<0.7~T_\mathrm{N}$ ($x=0.15$) and $1.9\text{~K}<T<0.7~T_\mathrm{N}$ ($x=0$ and $0.05$) were fit to a power law $C\propto T^\alpha$ (lines in the top panel of Fig.~\ref{HC}). The lower bound was imposed to avoid a Shottky-like contribution seen in all samples at the lowest temperatures, possibly due to nuclear spins.\footnote{In principle the nuclear contribution to specific heat could give additional information on the system. Unfortunately, due the unreasonably large time constants, the required detailed measurements at the lowest temperatures were not feasible.} The fitted exponent is plotted against Ge concentration in the bottom panel of Fig. 3. We see that with increasing $x$, the exponent rapidly crosses over from $\alpha\sim4.4$ in the parent material to $\alpha \sim 1.8$ in the disordered compounds.
Both these limiting values are distinct from expectations for a conventional antiferomagnet.  The latter has a linear spin wave dispersion. The result is a density of spin wave states which is quadratic with energy, and a Debye-like contribution to specific heat with $\alpha=3$. The observed large value of $\alpha$ in the disorder-free material is most likely related to the small anisotropy gap previously observed in these compounds.\cite{ZheludevKenzelmann2001} In contrast, a reduction of $\alpha$ with the introduction of disorder clearly indicates a proliferation of low-energy states compared to simplest spin wave model.

\subsection{$\mu$SR}

\subsubsection{Weak transversal field experiments}
 \begin{figure}
 \includegraphics[width=\columnwidth]{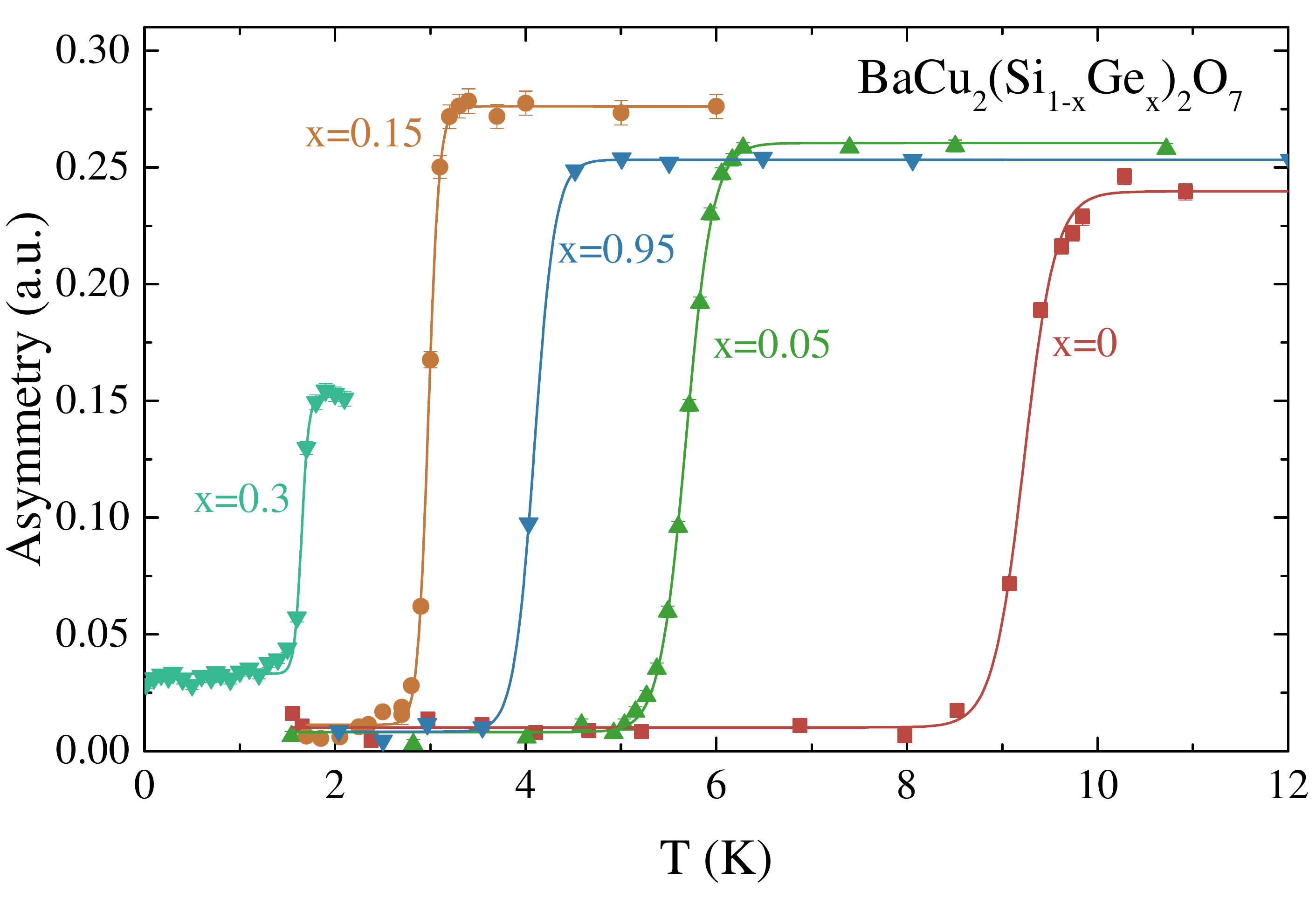}
 \caption{\label{wtf} (Color online) Measured temperature dependence of the asymmetry in weak transversal field, which is proportional to the magnetic volume fraction, for five different concentrations. The solid lines are fits to the data to determine the ordering temperature \TN \space and transition width.}
 \end{figure}
The \musr technique is used to accurately determine the magnetic volume fraction and the ordering temperature. The volume fraction is proportional to the initial asymmetry measured of the \musr spectra.\cite{YaouancReotier} For five of our Si-rich samples the measured the temperature dependence of the asymmetry is plotted in Fig. \ref{wtf}. Magnetic ordering is marked by a sharp step-like increase. At elevated temperatures, the sample is completely in the paramagnetic phase.  The observed variation of asymmetry in this regime is an instrument effect. For low temperatures, our results indicate that all samples are entirely ordered. For $x\leq 0.15$ this is immediately apparent in  the data collected on the Dolly and GPS instruments:  at low temperatures the asymmetry is almost zero. For the  $x=0.3$ we observe some residual asymmetry. This is, however, due to the about 20\% background in the LTF instrument, where these data were taken.

To determine the ordering temperatures, the data were analyzed using an empirical sigmoidal function:
\begin{eqnarray}
A = A_2+\frac{A_1-A_2}{1+\exp\left(\frac{T-T_\mathrm{N}}{\Delta T}\right)}
\end{eqnarray}
Here $A_1$, $A_2$ are the asymmetry below and above \TN \space respectively and $\Delta T$ is the apparent step width. Note that non-zero step widths are intrinsic to the technique. On the other hand, a particularly broad step would indicate a distribution of transition temperatures. The results of the fits are summarized in Tab. \ref{tab:ordering}, and plotted in Fig.~\ref{m0}. The excellent agreement with the calorimetric measurements is apparent. The step width is quite narrow and does not show any increase when $x$ deviates from zero or unity. All this testifies the quality and homogeneity of our samples.
\begin{table}[b]
\caption{\label{tab:ordering} Ordering temperature determined by specific heat and \musr \space measurements and apparent asymmetry step widths in transversal-field experiments.}
  \begin{ruledtabular}
    \begin{tabular}{cccc}
    $x$ & $T_\mathrm{N}$~(K), from $C(T)$ &$T_\mathrm{N}$~(K) from \musr & $\Delta T$~(K)\\
	\hline
    0     & 8.9(1)  & 9.23(1) & 0.163(10) \\
    0.05  & 5.4(1)   & 5.68(1) & 0.139(4) \\
    0.15  & 2.94(6)  & 2.97(1) & 0.057(3) \\
    0.3   & 1.63(4)  & 1.65(1) & 0.038(3) \\
    0.5   & 0.98(4)  & --    & -- \\
    0.95  & 4.0(1)  & 4.09(1) & 0.108(11) \\
    1     & 8.7(1)  &   --    & -- \\
    \end{tabular}%
  \end{ruledtabular}
\end{table}%

\subsubsection{Zero field \musr}
\label{subsubsec:zerofield}
\begin{figure}[h]
  \includegraphics[width=\columnwidth]{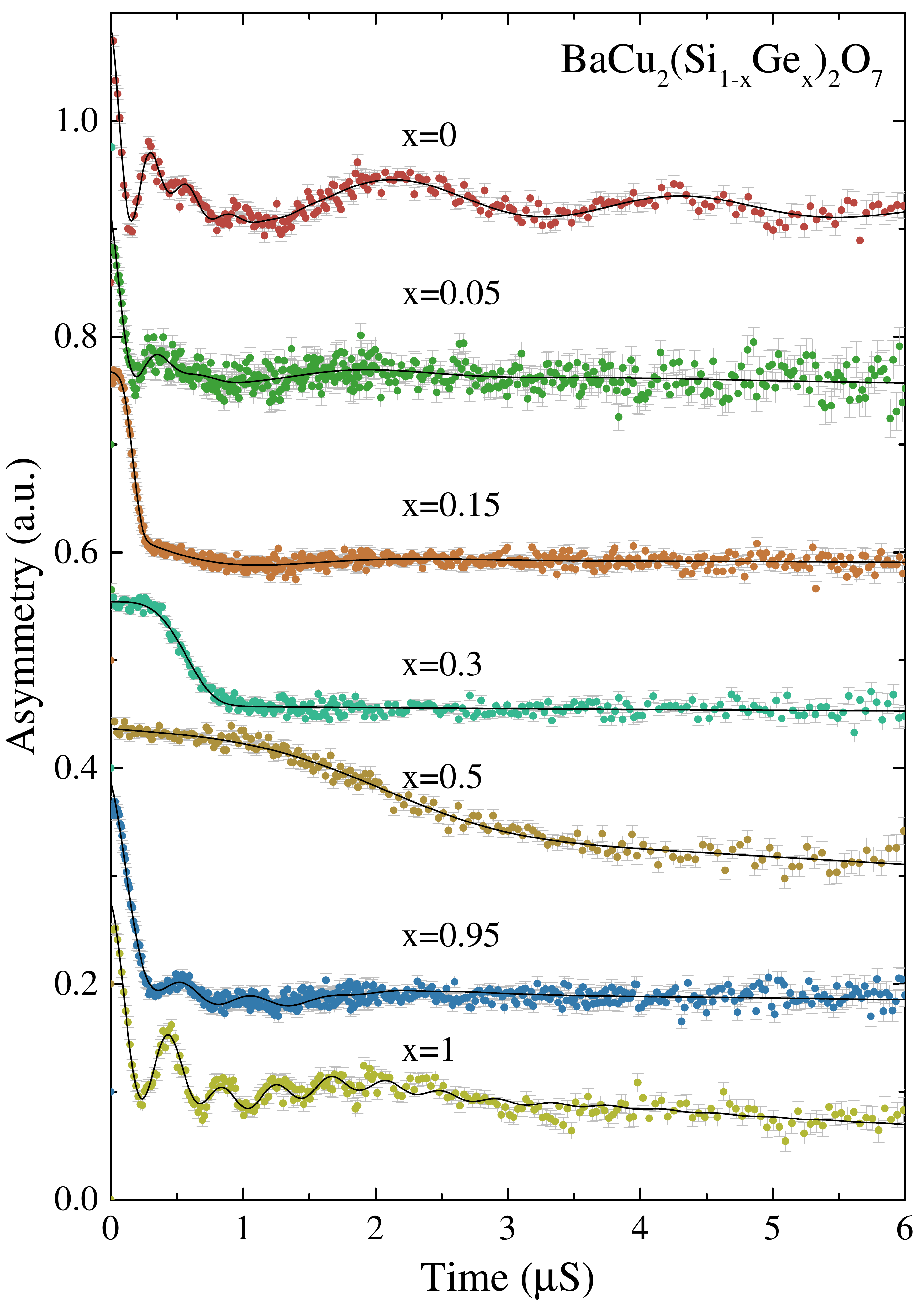}
 \caption{\label{spectra} (Color online) Zero field \musr time spectra of seven different concentrations. The data of  x=0.3 and x=0.5 were taken at 0.02 K and x=0, x=0.05, x=0.15, x=0.95 and x=1 at 1.6 K. The solid lines are fit to the data described in the text . From bottom to top the spectra are offset by 0, 0.1, 0.2, 0.4, 0.5, 0.7 and 0.85.}
 \end{figure}

The magnitude of the ordered magnetic moments was studied in zero field \musr experiments. Typical \musr spectra collected in a series of \BCSGO samples at the lowest accessible temperature  are shown in Fig. \ref{spectra}. Static internal fields due to magnetic order produce clear asymmetry oscillations, with at least two different frequencies in end compounds.  The component of the internal field that is parallel to the incoming muon spin causes a finite non oscillating asymmetry  at late times (``tail''). The immediate effect of disorder, {\it even at very low concentration} is the suppression of oscillations. The $x=0.05$ and $x=0.95$ materials still show at least two frequencies, but these are already heavily damped. At higher concentrations the spectra change to a more Gaussian-like shape, with the relaxation times decreasing towards $x=0.5$.
\\
 \begin{figure}[h]
 \includegraphics[width=\columnwidth]{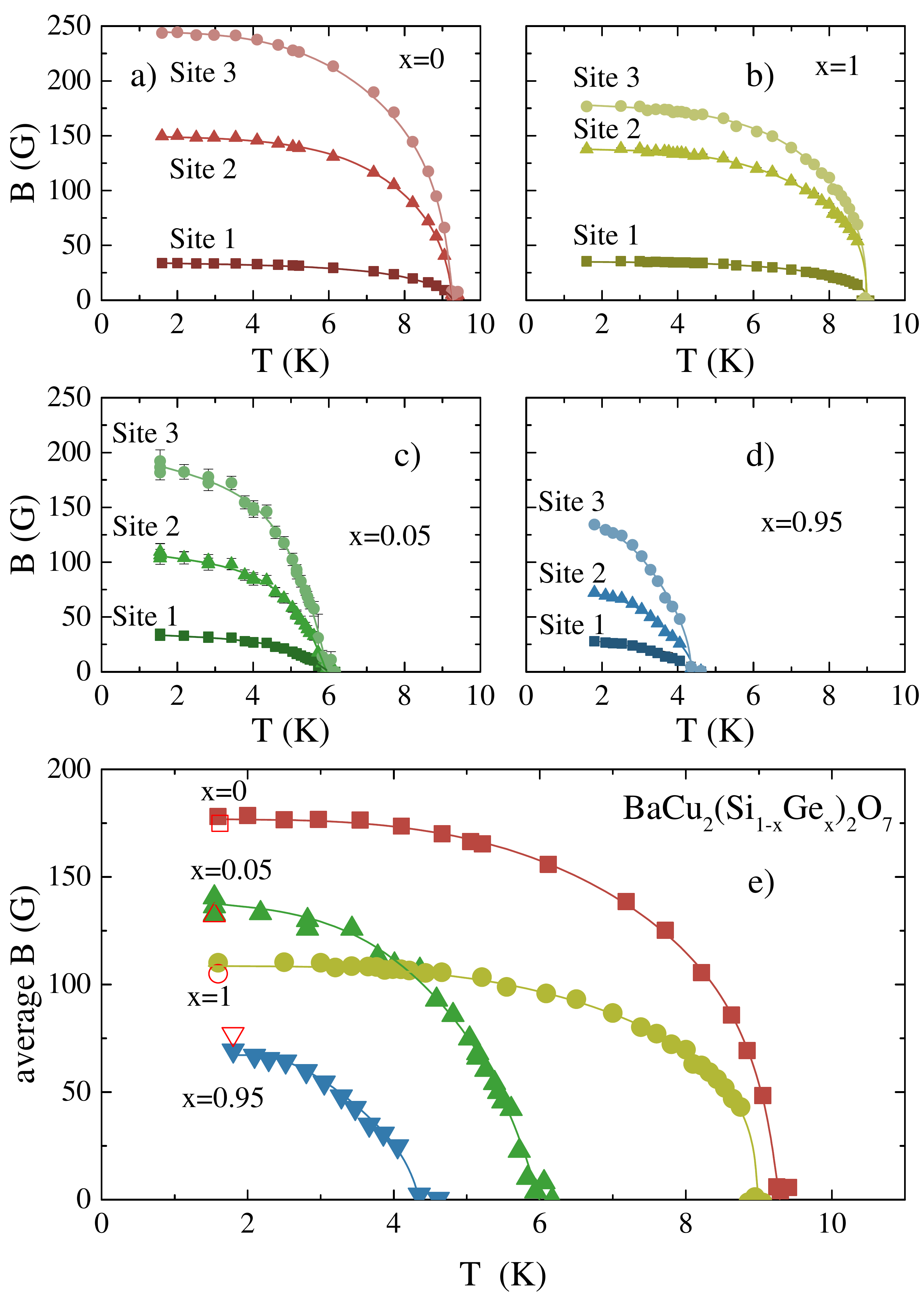}
 \caption{\label{field} (Color online) a-d) Measured temperature dependence of the internal fields at the three muon sites for a) $x=0$, b) $x=1$, c) $x=0.05$ and d) $x=0.95$  for the low-disorder samples (symbols), as described in the text. e) Temperature dependence of the average internal magnetic field, assumed to be proportional to the order parameter (solid symbols). The open symbols are the average fields determined by Fourier transformation.  In all cases, lines are guides for the eye.}
 \end{figure}
For a quantitative analysis, let us first focus on the regimes $x\geq0.95$ and $x\leq 0.05$, where damped oscillations are visible.
The pure material is known to order in a commensurate, nearly collinear antiferromagnetic structure.\cite{Tsukada1999} Thus, the multitude of observed precession frequencies is due to different muon stopping sites. An adequate description of the data can be obtained assuming three distinct sites. For this model, the asymmetry as a function of time is empirically given by:\cite{YaouancReotier}
\begin{eqnarray}
A&=&\frac 2 3 \left[A_1\cos(\omega_1t+\phi)\exp(-\lambda_1t) \right.\nonumber  \\
&+&A_2\cos(\omega_2t+\phi)\exp(-\lambda_2t) \nonumber  \\
&+&\left. A_3\cos(\omega_3t+\phi)\exp(-\lambda_3t)\right] \nonumber  \\
&+&\frac 1 3 (A_1+A_2+A_3)\exp(-\lambda_\mathrm{tail}t) \nonumber \\
&+& A_\mathrm{bg} \exp\left[-(\lambda_\mathrm{bg}t)^{\beta}\right]\label{A1}
\end{eqnarray}
The parameters $A_1$,$A_2$ and $A_3$ represent the muon fractions stopped at the different muon sites (site populations), respectively. Their ratio is not temperature dependent and was fit globally in each end-compound. For the disordered samples, the ratio was assumed to be the same as in the corresponding parent compounds. In Eq.~\ref{A1}, $A_\mathrm{bg}$ is the fraction of the muons which stop in a paramagnetic part of the sample, or possibly in the sample holder. Its value tends to be very small below $T_\mathrm{N}$. The corresponding parameters  $\lambda_\mathrm{bg}$ and $\beta$ were fit to spectra above $T_\mathrm{N}$ and kept fixed at low temperatures. $\lambda_\mathrm{tail}$ reflects the spin fluctuations of the muons with a polarization component parallel to the local magnetic field. We found it to be negligibly small and temperature independent. This implies that critical fluctuations are faster or slower than the accessible \musr time window.

The parameter $\phi$ tended to 0 in the initial fits, and was later fixed to $\phi=0$, as appropriate for a simple commensurate antiferromagnetic structure.
The parameters $\omega_1$, $\omega_2$ and $\omega_3$ are the mean precession frequencies at the three different muon sites in our model. They are related to the local internal magnetic fields on sites $i$ through $B_i=\gamma_{\mu} \omega_i$ with $\gamma_{\mu}=85.2 \times 10^{-3} \mathrm{rad s}^{-1}\mathrm{G}^{-1}$. The width of the corresponding field distributions are described by the relaxation rates $\lambda_1$, $\lambda_2$ and $\lambda_3$. The latter were found to vary only slightly below $T_\mathrm{N}$. Upon introduction of disorder the relaxation rates increase considerably. Assuming that all the internal fields are proportional to the static ordered magnetic moment,  the ratios $B_i:B_j$ were treated as temperature-independent parameters in each sample. The temperature dependence of the three internal fields for $x=0$, 0.05, 0.95 and 1 are shown in Fig. \ref{field} (a-d). They show typical order-parameter behavior. Within each sample, the saturation fields at the three sites are rather different. Their ratios are concentration-dependent. Fig. \ref{field} (e) shows the temperature evolutions of the mean site-fields. These were calculated taking the respective populations of the three  muon sites into account. One immediately sees that the saturation mean field and the ordering temperature decrease with disorder.

\begin{figure}[h]
 \includegraphics[width=\columnwidth]{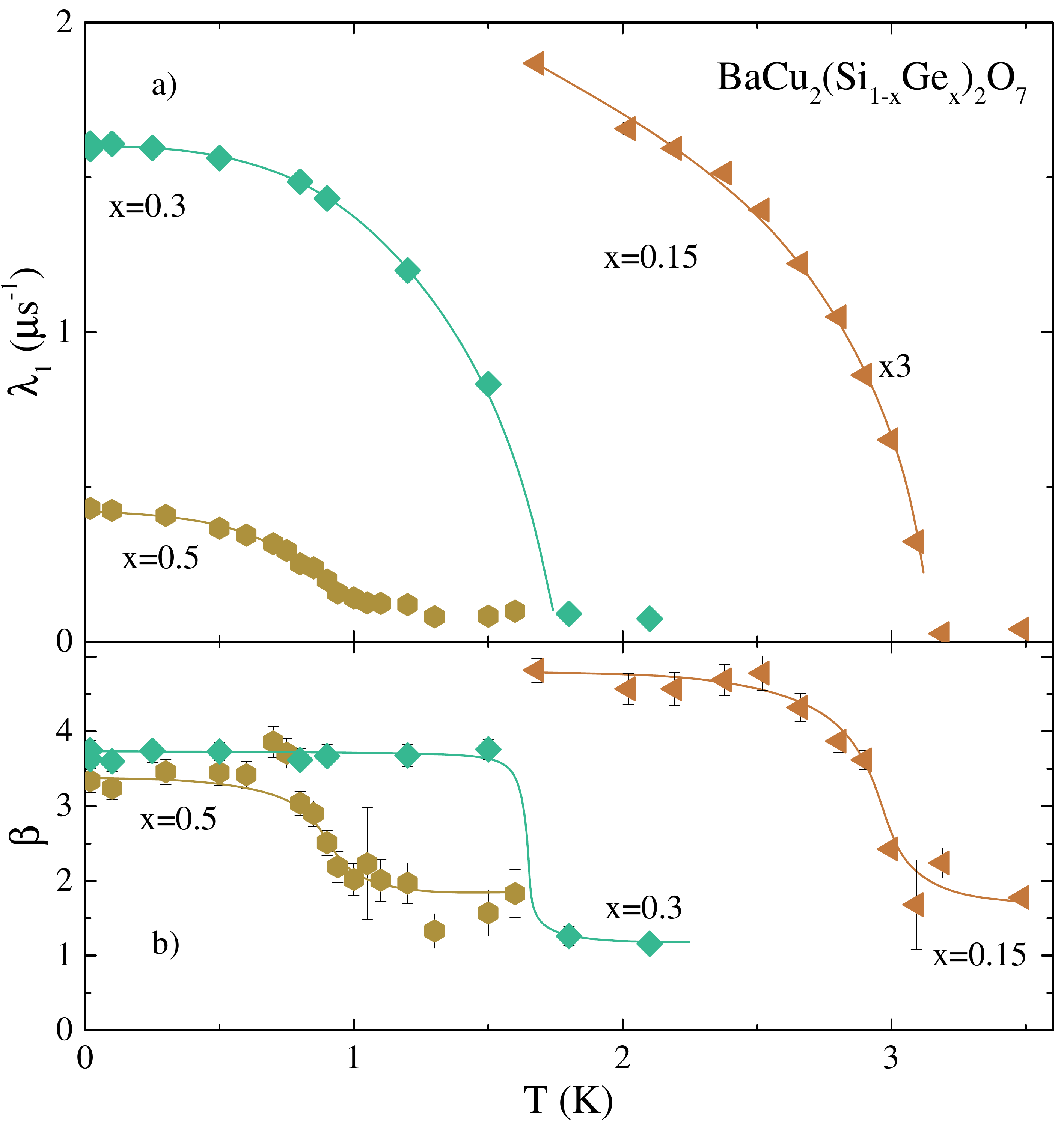}
 \caption{\label{lambda} (Color online) Temperature dependence of the fitting parameters $\lambda$ (a) and $\beta$ (b) for highly disordered samples (symbols), as described in the text. Lines are guides for the eye.}
 \end{figure}

For the $x=0.15$ sample, the functional shape of the \musr ZF-spectra changes and the three-site model is not applicable any more. As seen in Fig.~\ref{spectra}, the spectrum at this concentration shows a fast decay at early times, followed by a slow and weak oscillation. Denoting the oscillating fraction as A$_1$, and that with the fast decay as A$_2$, for this concentration we used the following empirical fitting function:
\begin{eqnarray}
\label{x0p15}
 A&=&\frac 2 3 \left[A_{1} \cos(\omega_1t+\phi) \exp(-\lambda_1t) +  A_{2} \exp\left[-(\lambda_2t)^{\beta}\right]\right]+  \nonumber  \\
  &+& \frac 1 3 (A_1 + A_2) \exp(-\lambda_{\mathrm{tail}}t).
\end{eqnarray}
In these measurements, the background contribution was negligible. The ratio $A_1:A_2$ was fit globally with the resulting value of 1:1.336(4). In Eq.~\ref{x0p15}, $\omega_1$ is the frequency of oscillations. The phase $\phi$ was a global fit parameter for all temperatures yielding $\phi=27(1)^\circ$. The parameter $\lambda_1$ is the damping rate of the oscillations.  $\lambda_{tail}$ is temperature independent and negligibly small as in the previously described concentrations. In Eq.~\ref{x0p15} the second term represents all muons that stop in rather different magnetically environments. The stretched exponential is an attempt to account for the corresponding broad distribution of internal magnetic fields, which leads to a superposition of multiple harmonic precession. The temperature dependency of the parameters $\beta$ and $\lambda_2$ are shown in Fig. \ref{lambda}. Upon lowering the temperature, a strong increase of $\lambda_2$  is observed at the transition point. Simultaneously, $\beta$ changes in a step-like fashion from $\beta \sim1.9$, to $\beta \sim4.6$. The high-temperature value is in agreement with our expectation for the paramagnetic phase,\cite{YaouancReotier} while the high value at low temperature signals a broad unconvential distribution of local fields. The oscillation frequency $\omega_1$ increases below the ordering temperature, as expected.

For stronger disorder, in samples with $x=0.3$ and $x=0.5$, the oscillatory contribution becomes indiscernible. The corresponding data were analyzed using only the stretched exponential function:
\begin{eqnarray}
 A&=&\frac 2 3 A_1\exp\left[-(\lambda_1t)^\beta\right] +\frac 1 3 A_1\exp(-\lambda_\mathrm{tail}t)+ \nonumber \\
 & +& A_\mathrm{bg}\exp(-\lambda_\mathrm{bg}t).
\end{eqnarray}
As for $x=0.15$, the stretch exponential describes the superposition of a distribution  of oscillating frequencies. The second term is again related to the muon spin component parallel to the internal field. The non-negligible background contribution is due to muons stopped in the $^3$He-dilution refrigerator used in this run. A global fit for each sample gives $A_\mathrm{bg}=0.122(3)$ and $A_\mathrm{bg}=0.106(1)$, for $x=0.5$ and $x=0.3$. The temperature dependence of $\beta$ and $\lambda_1$ are shown in Fig. \ref{lambda}. The exponent increases upon cooling through $T_\mathrm{N}$. Its saturation values at the lowest temperature are rather similar: $\beta\approx3.6$ for $x=0.3$ and $\beta\approx3.4$ for $x=0.5$, respectively. In contrast, the relaxation rates  $\lambda_1$ are about 4 times shorter for $x=0.5$, as compared to $x=0.3$.

\subsubsection{Fourier inversion}
  \begin{figure}[!]
  \includegraphics[width=\columnwidth]{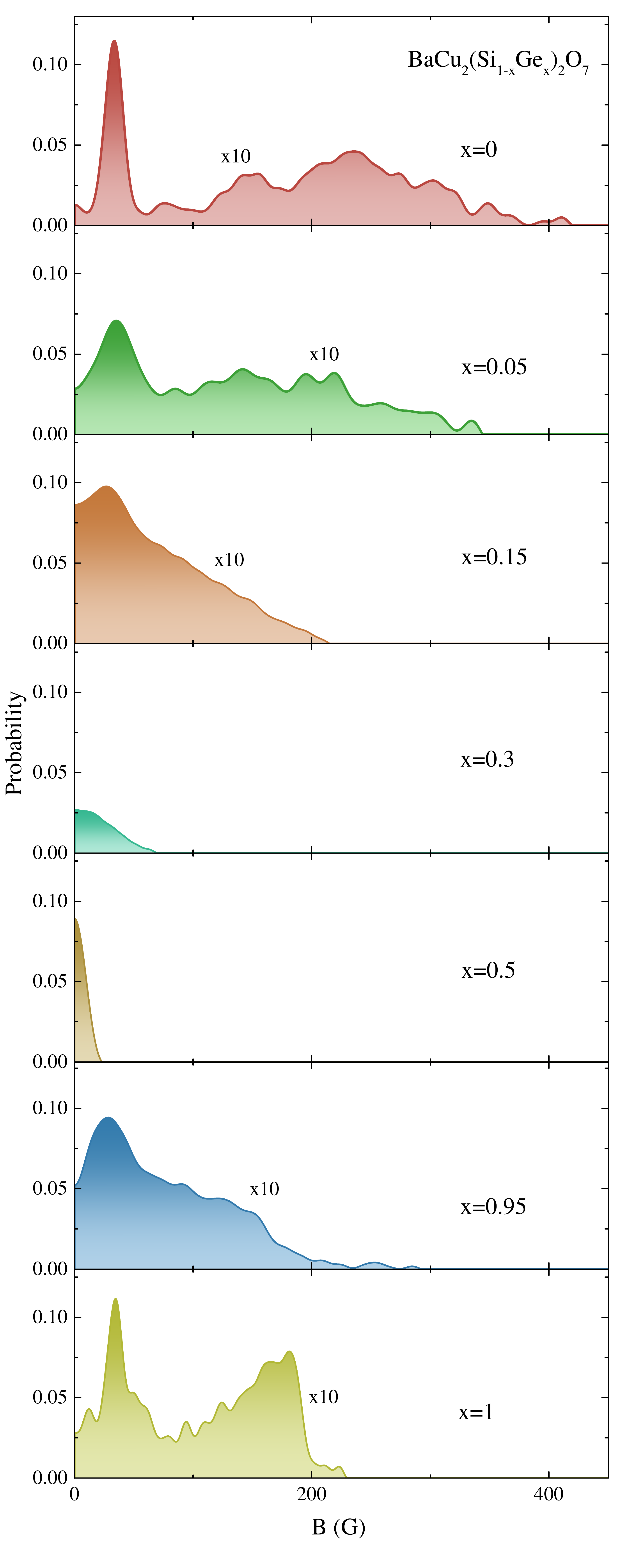}
  \caption{\label{fourier} (Color online) Measured internal magnetic field distribution in \BCSGO. The data of  $x=0.3$ and $x=0.5$ were taken at $T=0.02$~K.  For $x=0$, $x=0.05$, $x=0.15$, $x=0.95$ and $x=1$ $T=1.6$~K.}
  \end{figure}
 Additional information was obtained using an alternative, model-independent method to analyze the zero-field \musr data. It is based on the fact that
 the  muon time-spectra are temporal Fourier transforms of the probability distribution of the muon spin Larmor precession frequencies in the sample.\cite{YaouancReotier}
 Thus, a Fourier transform of the measured spectra will yield the probability distribution of the {\it modulus} of internal fields. Before applying this procedure to our data, we subtracted the ``tail'' and background contributions determined in model fits, as described above. The result of the subsequent Fourier inversion is shown in Fig. \ref{fourier}. Both parent materials show two well pronounced peaks. A third maximum can be identified as a shoulder of the higher-field peak. The central positions of these features correspond to the fitted field values in our model-based data analysis. At 5\% disorder on both ends we see a broadening of all features and a shift to smaller fields. At further higher concentrations the peaks merge and the probability weight builds up at very small fields. For $x=0.3$ and $x=0.5$ the probability maxima are actually at zero field.

\section{Discussion}
It is clear from the above that chemical disorder in \BCSGO does not eliminate the well-defined magnetic ordering transitions at low temperatures. On the other hand, it has two obvious and rather drastic effects on the ordered state. First, disorder tends to  suppress the average spontaneous static magnetization and the ordering temperature. Second, it results in a broad {\it distribution} of magnitudes of local fields (static moments) in the sample. We shall address these two aspects of the observed behavior separately.

\subsection{Average ordered moment}
Despite the fact that several muon sites are involved, we shall make the assumption that the average ordered moment is proportional to the average local field seen by the muons. Implicitly, we are assuming that the muon stopping sites are the same across the Ge concentration range. In application to our data, the mean local field is readily obtained from the measured probability distribution shown in Fig. \ref{fourier}. For $x=0$, a ,a moment of $m_0=0.108\mu_b$ can be estimated from chain-Mean-Field (MF) theory, using the known values of $J$ and $T_\mathrm{N}$ .\cite{Tsukada1999,Schulz1986}  Note the strong suppression of $m_0$ by quantum fluctuations enhanced by the system's one-dimensionality. Using the chain-MF estimate as a ``calibration point'' for the measured mean local fields in other samples, in Fig.~\ref{m0} we plot the saturation moment as a function of Ge concentration. Note the logarithmic $y$-axis in this graph. As a consistency check, for the pure Ge sample, our procedure yields $m_0=0.072\mu_b$. This is only about 12\% off the chain-MF estimate $m_0=0.064$. One can thus hope that a 20\% accuracy of our method holds for the entire Ge-concentration range.

Our analysis reveals a drastic suppression of the ordered moment already at very low levels of chemical randomness.  This behavior emphasizes the quantum nature of the spin chains involved. In contrast, a classical antiferromagnet with fully saturated N\'{e}el sublattices would not be affected by bond disorder at all. Its ground state would remain the fully polarized two-sublattice state. In our case though, on the Si-rich end the decrease of $m_0$ is {\it exponential}. A similar (but somewhat less drastic) behavior was observed in the organic weakly coupled spin chain material Cu(py)$_{2}$Cl$_{1-x}$Br$_{x}$ (CPX)\cite{Thede2012}. At the same time, these trends totally contradict chain-MF calculations for bond-disordered chains, where both $T_\mathrm{N}$ and $m_0$ were predicted to increase.\cite{Joshi2003}
\begin{figure}[!]
  \includegraphics[width=\columnwidth]{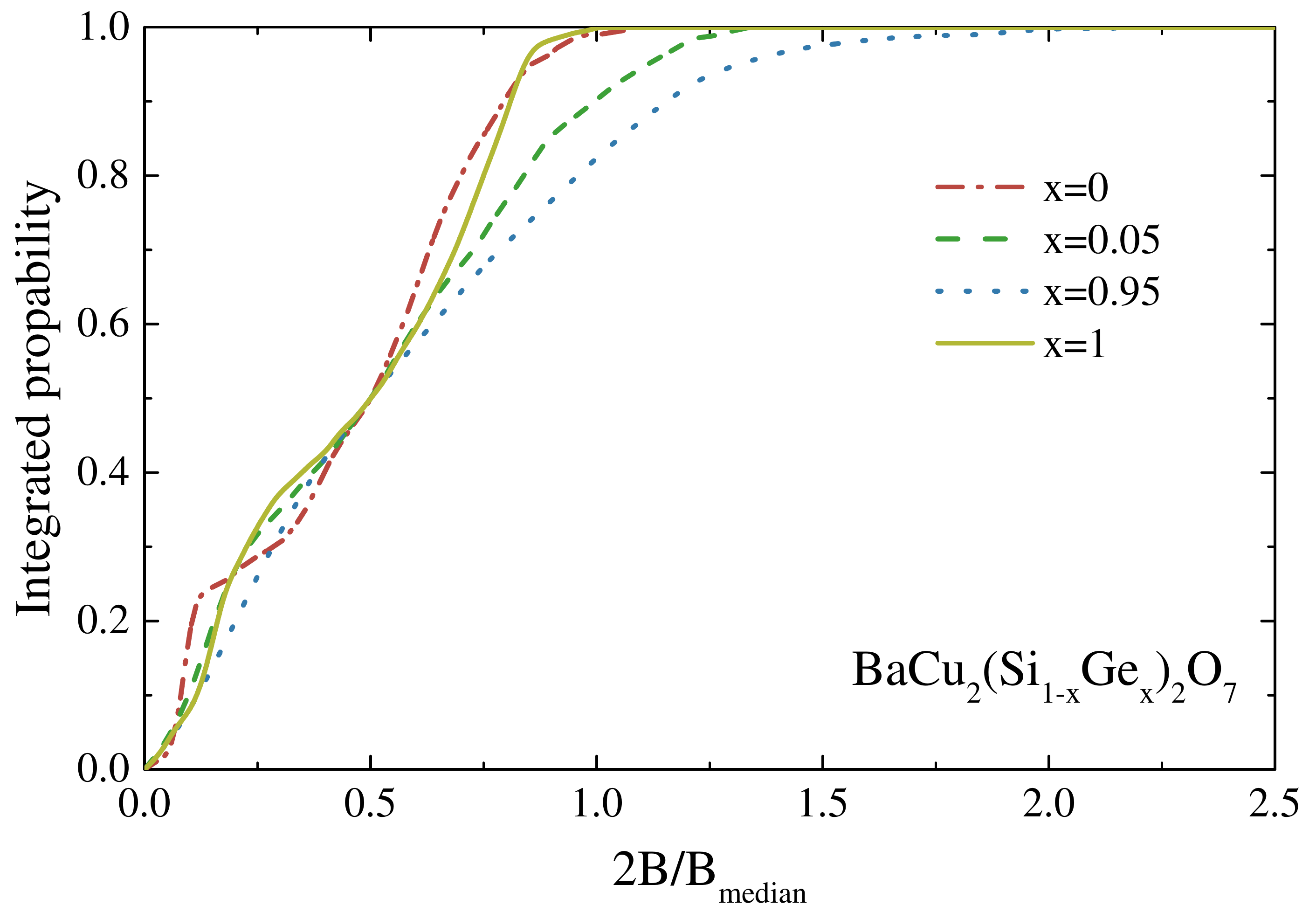}
  \caption{\label{integr} (Color online) Measured cumulative probability distribution of local fields in low-disorder \BCSGO samples. The $x$ axis is normalized to the median value.}
  \end{figure}

\subsection{Distribution of ordered moment}
The most important result of this work is the study of the {\it distribution} of ordered moments in samples with bond randomness. This kind of analysis was not possible for our previous experiments on CPX.\cite{Thede2012}  In the case of \BCSGO, already the fact that the data are described by a stretched exponential relaxation with a large exponent $\beta$ indicates a very unconventional distribution of local fields. In fact, a substantial broadening of the field distribution is already present at very small disorder levels. The most unbiased way to visualize this, is by integrating the measured probability distributions, as shown in Fig. \ref{integr}. Here the scale of the $x$ axis for each material is chosen so as to have their median field points coincide. We see that the magnitudes of local fields in both weakly disordered samples with $x=0.05$ and $x=0.95$ have a much broader relative distribution than in the parent material. In particular, the proportion of relatively large fields is substantially increased by disorder.

As mentioned above, the ordered moment in weakly coupled spin chains is suppressed by quantum spin fluctuations. Any bond strength disorder will modulate the strength of these quantum fluctuations in a spatially random manner. The result will be an inhomogeneous spatial distribution of ordered moments. However, it is clear that at least in our low-disorder samples, the inhomogeneity is not confined to the rare and isolated substitution sites. Indeed, for $x=0.05$ and $x=0.95$, the vast majority of muons rest in a relatively homogeneous environment, similar to that in the corresponding parent compound. Rather, the broad distribution, as well as a disproportionate prevalence of relatively large local fields (moments), is due to the  RS nature of the individual chains.

Consider a single random-bond chain. In the ordered phase, at the MF level, it is subject to a staggered field generated by neighbouring chains. The RS ground state can be viewed as a collection of {\it non-interacting} effective spin dimers.\cite{Ma1979,Dasgupta1980,Fisher1994} The probability distribution of the corresponding dimer coupling strengths $J$ is peaked at zero, i.e., there is an abundance of very weak effective dimers, that also will tend to involve physically distant spins.\cite{Ma1979,Dasgupta1980,Fisher1994} Each such dimer will get polarized by the staggered mean field. The resulting induced moment will be proportional to the dimer's susceptibility, which, at low temperatures, is inversely proportional to $J$. Thus, the distribution of magnetic moments will echo the distribution of inverse exchange constants. It will be very broad, and will have a disproportionate number of large moments. These are due to the particularly weak dimers in the RS state. Unfortunately, to date, neither the probability distribution nor the spatial distribution of ordered moments in weakly coupled RS chains have been addressed theoretically or numerically.

The inhomogeneous ground state will undoubtedly affect the low-energy excitations. A hint of this phenomenon is present in our calorimetry data. In the presence of disorder, quasimomentum ceases to be a good quantum number and the entire spin wave excitation picture breaks down. The observed decrease in the effective $C\propto T^\alpha$ power law exponent signals a disproportional abundance of low-energy excitations.  A decrease of specific heat is directly related to a modification. At this point we can only speculate that this is a consequence of the RS nature of the individual chains too. Undoubtedly more insight will come from further theoretical studies.

\section{Conclusion}
In summary, even weak bond strength disorder strongly affects magnetic ordering in weakly coupled quantum spin chains. The average ordered moment is drastically suppressed. At the same time the distribution of local static moments is highly inhomogeneous, with a disproportionate number of relatively large moments. The thermodynamics of the ordered state is also affected. All this appears to be a direct consequence of ``infinite disorder'' in individual random-bond chains.

\bibliography{azbib}
\end{document}